# Spin – Phonon Coupling in Nickel Oxide Determined from Ultraviolet Raman Spectroscopy

E. Aytan[1,2], B. Debnath[3], F. Kargar[1,2], Y. Barlas[3], M. M. Lacerda[2,4], J. X. Li[5], R. K. Lake[3,6], J. Shi[5,6] and A.A. Balandin[1,6,*]

[1]Nano-Device Laboratory (NDL), Department of Electrical and Computer Engineering, University of California, Riverside, CA 92521 USA

[2]Phonon Optimized Engineered Materials (POEM) Center, Materials Science and Engineering Program, University of California, Riverside, CA 92521 USA

[3]Laboratory for Terascale and Terahertz Electronics (LATTE), Department of Electrical and Computer Engineering, University of California, Riverside, CA 92521 USA

[4]Campus Duque de Caxias, Universidade Federal do Rio de Janeiro, RJ, 25245-390, Brasil

[5]Department of Physics and Astronomy, University of California, Riverside, CA 92521 USA

[6]Spins and Heat in Nanoscale Electronic Systems (SHINES) Center, University of California, Riverside, CA 92521 USA

---

[*] Corresponding author (AAB): E-mail: balandin@ee.ucr.edu ; Web-page: http://balandingroup.ucr.edu/

**Nickel oxide (NiO) has been studied extensively for various applications ranging from electrochemistry to solar cells[1,2]. In recent years, NiO attracted much attention as an antiferromagnetic (AF) insulator material for spintronic devices[3–10]. Understanding the spin – phonon coupling in NiO is a key to its functionalization, and enabling AF spintronics' promise of ultra-high-speed and low-power dissipation[11,12]. However, despite its status as an exemplary AF insulator and a benchmark material for the study of correlated electron systems, little is known about the spin – phonon interaction, and the associated energy dissipation channel, in NiO. In addition, there is a long-standing controversy over the large discrepancies between the experimental and theoretical values for the electron, phonon, and magnon energies in NiO[13–23]. This gap in knowledge is explained by NiO optical selection rules, high Néel temperature and dominance of the magnon band in the visible Raman spectrum, which precludes a conventional approach for investigating such interaction. Here we show that by using ultraviolet (UV) Raman spectroscopy one can extract the spin – phonon coupling coefficients in NiO. We established that unlike in other materials, the spins of Ni atoms interact more strongly with the longitudinal optical (LO) phonons than with the transverse optical (TO) phonons, and produce opposite effects on the phonon energies. The peculiarities of the spin – phonon coupling are consistent with the trends given by density functional theory calculations. The obtained results shed light on the nature of the spin – phonon coupling in AF insulators and may help in developing innovative spintronic devices.**

The use of spin currents in spintronic devices instead of charge currents in electronic devices is advantageous for ultra-low energy dissipation[11,12]. NiO occupies a special role among spintronic AF insulator materials[4–9]. Its high Néel temperature, $T_N$=523 K, allows for device operation in the AF state at room temperature (RT) and above. Demonstrated coherent THz control of AF spin waves suggest a possibility of ultra-high speed operation of NiO spintronic devices. NiO has been used for spin-fluctuation mediated transmission of spin current[3]. Unlike in ferromagnetic (FM) materials, an external magnetic field does not cause significant change in the static spin structure or spin wave excitations in NiO[24], which is a major benefit for spintronic memory and logic applications. NiO is a strongly correlated material and electrical insulator in both its AF and paramagnetic (PM) phases. Interaction between Ni 3d-valence electrons is agreed to be the reason for the insulating character below and above $T_N$. Despite many years of experimental and theoretical studies, there is significant discrepancy in measured and calculated electronic, phonon and magnon band structures in NiO[13–23]. It has been suggested that the spin – phonon interaction is likely behind this discrepancy, although no direct proof has been offered. The strength of this interaction is not known and its specifics are not understood. The importance of the knowledge of the spin – phonon coupling in NiO goes beyond fundamental science owing to its relevance to magnon damping in spin wave devices.

The strength of the spin – phonon coupling in many FM and AF materials, e.g. $FeF_2$, $MnF_2$, $NiF_2$, has been determined from the spectral position of the phonon peaks in the Raman spectra taken over wide temperature ranges extending below and above the Néel temperature $T_N$ [25–30]. The deviation of the phonon peak position below $T_N$ from the theoretically predicted phonon anharmonic decay curve was attributed to the effects of spin texture on the phonon energies[25–30]. The anharmonic decay dependence was obtained from theory fitting to experimental data points above $T_N$. This conventional method does not work for NiO for several specific reasons. The first-order longitudinal optical (LO) and transverse optical (TO) phonons are not Raman active in rock-salt crystals[31]. A weak rhombohedral lattice distortion in NiO below $T_N$ or the presence of defects can make LO and TO phonons visible in the Raman spectrum but their low intensity does not allow for the extraction of the coupling strength. The second-order LO and TO bands, denoted as 2LO and 2TO, have much higher intensity but always appear within the large background of the second-order two-magnon (2M) band, which does not permit an accurate phonon peak position analysis[13,14,19,32,33]. Moreover, because the Néel temperature in NiO is very high ($T_N$=523 K), there is a limited temperature range above $T_N$ available for fitting to the anharmonic curve, before irreversible structural changes start to happen to the NiO crystal lattice[23]. All of these factors together make the study of spin – phonon interactions in NiO challenging.

We employed UV Raman spectroscopy, which allows for the suppression of the 2M band, to overcome the above mentioned difficulties, and determine the spin coupling with LO and TO phonons. The single crystal NiO slab samples, selected for this study, have the crystalline growth plane oriented in the [111] direction. The X-ray diffraction (XRD) inspection confirmed their high quality. NiO has a two-sub-lattice AF crystalline structure consisting of ferromagnetically (FM) aligned (111) planes that are AF aligned with respect to each other[4,8,9]. The AF ordering of NiO is due to the super-exchange interaction mediated by O orbitals[32]. Above $T_N$, NiO has a face-centered cubic (FCC) structure with PM ordering while below $T_N$, NiO lattice is characterized by a weak rhombohedral distortion[34]. A schematic of the crystal structure and representative XRD data are shown in *Supplementary Figure S1*.

Raman spectroscopy (Renishaw InVia) was performed in the backscattering configuration under visible (2.54 eV or 488 nm) and UV (3.81 eV or 325 nm) laser excitation. The excitation laser power was limited to 2 mW to avoid any possible effects on the sample surface due to local heating. Spectra at different temperatures were measured with the sample in a cold-hot cell with a constant flow of pure Ar gas to avoid surface oxidation. The temperature, *T*, was accurately controlled at each step. The sample was heated or cooled to a certain *T* and then maintained at that *T* for ~10 minutes to ensure that the entire sample reached a constant *T*. Figure 1 (a-b) shows room temperature (RT) Raman spectrum of NiO under visible and UV excitation. The weak peaks observed in the range from 350 cm$^{-1}$ to 410 cm$^{-1}$ and from 520 cm$^{-1}$ to 580 cm$^{-1}$ are the TO and LO phonon modes, respectively. The stronger second-order peaks are assigned to 2TO (~738 cm$^{-1}$), 2LO (1142 cm$^{-1}$) and a combination of TO+LO (913 cm$^{-1}$) phonon modes, which is consistent with the literature[13,19,35]. The spectral positions are given for UV Raman spectra, which may deviate from those in visible Raman spectra due to the difference in the probing phonon wave vector $k_p=4\pi f n/c$, and possible resonance excitation effects for the compositional peaks (here *f* is the excitation frequency, *c* is the speed of light and *n* is the refractive index of the material at given frequency). The probing phonon wave vector for UV laser excitation is $k_p=0.124$ nm$^{-1}$, which is approximately twice of that for the visible. The second-order 2M peak around 1500 cm$^{-1}$ is a dominant feature of the visible Raman spectrum. It originates from two Brillouin zone (BZ) edge magnons, likely *X* or *Z* points, propagating with opposite momenta[17,36]. It is interesting to note that there are always substantial deviations in the measured and theoretically predicted energies of the 2M band, likely related to the spin – phonon coupling. The most striking feature of the UV Raman spectrum is absence of 2M band.

The unusually strong difference in the Raman spectra of NiO under visible and UV excitation can be explained by the following considerations. The UV excitation energy that we selected (3.81 eV) is close to the NiO optical band gap energy $E_G$=3.80 eV measured by the reflectance studies[37,38]. One should note that the exact $E_G$ value is a non-trivial question for NiO since the reported energies, obtained by different experimental techniques, vary from 3.5 eV to 4.3 eV[37–39]. Matching the resonant band-to-band transition increases the light absorption while simultaneously enhancing the phonon Raman signatures near the $\Gamma$ point. The increased light absorption leads to decreased penetration depth and substantially reduced interaction volume. Unlike $\Gamma$-point phonons, the BZ-edge 2M band is non-resonant due to a difference in the energy scale[17,36,40]. As a result, the 2M band almost completely disappears in UV Raman spectrum while 2LO and 2TO bands, resonantly enhanced, remain virtually the same or even enhanced despite the reduction in the interaction volume. It is the absence of the strong 2M background which allowed us to accurately determine the 2LO and 2TO band spectral positions and their temperature dependence (see *Supplementary Figure S2*).

To elucidate the spin-phonon interaction in NiO, we conducted UV Raman spectroscopy for temperatures below and above $T_N$. Figure 1 (c-d) shows the evolution of 2TO and 2LO peaks over a wide temperature range. We were careful to keep the temperature below $T_M$~700 K to avoid any irreversible structural changes with the material[23]. Below $T_N$, the phonon peak position is affected by the AF spin texture[41]. Thus, in order to determine the strength of the spin – phonon coupling, one needs to separate it from the baseline temperature dependence due to the crystal lattice anharmonicity. The anharmonic dependence of the phonon frequency is given within second-order perturbation theory as[42,43], $\omega_A(T) = \omega(0) - A\left(1 + 2/(e^{\frac{\hbar\omega(0)}{2K_BT}} - 1)\right)$, where $A$ and $\omega(0)$ are the fitting constants, while $\omega_A(T)$ is the phonon frequency determined by the anharmonicity alone. The expression for the phonon anharmonic temperature dependence assumes an optical phonon decay to two BZ-edge acoustic phonon with opposite momenta. The experimental data points at temperatures $T > T_N$, where NiO becomes PM, but lower than $T_M$, are those that should be used for fitting with the goal of determining the anharmonic phonon decay baseline. One can see that there is a significant deviation of the experimental data points below $T_N$ from this dependence, indicating a strong spin – phonon coupling in NiO.

[Figure 1: Experiment]

The renormalized phonon frequency, $\omega_R$, due to the spin-dependent effects is given as[25,27,44] $\omega_R = \omega_0 + \lambda \langle S_i \cdot S_j \rangle$. Here, $\omega_0$ denotes the phonon frequency in the absence of spin correlations, and $\langle S_i \cdot S_j \rangle$ is the spin-spin correlation function of the adjacent spins. For $T > T_N$, the spin correlation function approaches to zero, terminating any spin-phonon effect in paramagnetic phase, i.e. $\omega_R \to \omega_0$. Hence, the spin-phonon coupling constant $\lambda$ is proportional to the difference in the frequency between the two phases, $\Delta\omega = (\omega_R - \omega_0)$. From this deviation, we calculated the spin-phonon coupling factors, $\lambda$, for TO and LO phonons as -7.9 cm$^{-1}$ and 14.1 cm$^{-1}$, repectively. The AF spin texture in NiO results in substantial softening of the TO phonon and hardening of the LO phonon. For comparison, the spin-phonon coupling in NiO is substantially stronger than that in other common AF materials such as MnF$_2$ ($\lambda$=0.4 cm$^{-1}$)[25], FeF$_2$ ($\lambda$=1.3 cm$^{-1}$)[25] or ZnCr$_2$O$_4$ ($\lambda$=3.2 cm$^{-1}$ – 6.2 cm$^{-1}$)[44] although not as strong as in NaOsO$_3$ ($\lambda$~40 cm$^{-1}$)[45] and CuO ($\lambda$~50 cm$^{-1}$)[46], which is perhaps the strongest reported to date. In order to rationalize the obtained experimental results and understand if the strong spin-phonon coupling can be behind the observed discrepancy in the experimental and theoretical data for the phonon energies, we calculated the full phonon dispersion of NiO using *ab initio* density functional theory (DFT). Details of the $\lambda$ extraction and DFT calculations are provided in the *Methods*.

Figures 2 (a) and (b) show the phonon dispersion of AF NiO and non-magnetic NiO (*i.e.* no spin ordering), respectively. The calculations were performed for the primitive cell containing two Ni atoms and two O atoms. The corresponding BZ is shown in Figure 2 (c). The *x*-axis labels in Figure 2 (a-b) are in the basis of the reciprocal lattice vectors of the AF primitive cell shown in Figure 2 (c). We identify the first-order TO and LO modes from the corresponding atomic displacements (see *Supplementary Figure S3 and Supplementary Movie*). The positions of the experimental Raman peaks are indicated by the spheres. At the Γ point, the calculated frequencies of the slightly split TO mode are 360 cm$^{-1}$ and 367 cm$^{-1}$. The calculated TO mode splitting at the Γ point of 7 cm$^{-1}$ depends on the BZ direction, and it is close to literature values[41,47]. The calculated frequencies of the LO' and LO modes are at 535 cm$^{-1}$ and 571 cm$^{-1}$, respectively. Three of the calculated Γ-point phonon energies agree well with the UV Raman peaks, while the TO' Raman peak at 413 cm$^{-1}$ has no corresponding calculated energy at Γ. A comparison of the calculated LO and TO freuquencies in Figures 2 (a) and (b) shows that the inclusion of AF spin texture hardens the LO phonon (increases the frequency) and softens the TO phonon (decreases the frequency). The calculated relative shift, Δω, is -60.5 cm$^{-1}$ and 146.2 cm$^{-1}$ for TO and LO phonons, respectively. Hence, we expect that the spin phonon coupling of LO phonon will be, approximately, two times larger in magnitude than that of the TO phonon, and the signs of the shifts will be opposite, specifically negative for the TO phonon and positive for the LO phonon. These trends in relative magnitudes and signs are consistent with the experimental results (see Figure 1 (c-d)). Our experimental results and ab initio calculations prove that inclusion of the spin – phonon interactions is essential for an accurate description of the phonon frequencies in NiO.

[Figure 2: Theory]

The experimental TO' peak is absent in the DFT spectrum shown in Figure 2 (a), however its source is indicated by the arrow. To better understand this feature, we double the size of the primitive cell in Figure 2 (c) to obtain the unit cell and corresponding BZ shown in Figure 2 (d). Performing the same phonon dispersion calculation for AF NiO with this unit cell gives the phonon dispersion shown in Figure 2 (e). The *x*-axis labels are in the basis of the reciprocal lattice vectors of the BZ in Figure 2 (d). We now observe that the experimental TO' mode aligns with the calculated degenerate mode at Γ. To analyze this energy, we write the reciprocal lattice vectors of the larger unit cell in terms of those of the primitive unit cell to obtain $g'_a = \frac{1}{2}(g_a + g_b)$, $g'_b =$ $\frac{1}{2}(g_a + g_c)$ and $g'_c = \frac{1}{2}(g_b + g_c)$. The *k*-point denoted as [0.5,0,0.5] $= \frac{1}{2}(g_a + g_c)$ in Figure 2 (a-b) lies at the center of the small face of the BZ in Figure 2 (c), and it is equal to $g'_b$ in Figure 2 (d). Since the reciprocal lattice vectors correspond to equivalent Γ points, this *k*-point appears at Γ in Figure 2 (e). Thus, the calculated modes at the TO' frequency originate from the center of the six small faces of the BZ of the AF primitive cell. An analysis of the atomic displacements of this mode, shows that they are similar atomic to the displacements of the TO peak (see *Supplemental Video*).

In conclusion, we succeeded in determining the spin-phonon coupling strength in NiO by using UV Raman spectroscopy and focusing on the second-order phonon bands. We undertook an unconventional approach in order to avoid a dominant 2M band in the visible Raman spectrum and mitigate the problem of prohibitive optical selection rules for the first-order phonons in NiO. It was found that the AF spin texture in NiO produces strong and opposite effects on the LO and TO phonon energies. *Ab initio* calculations confirm that inclusion of spin texture is essential for obtaining the correct phonon energies and explaining the discrepancy in the theoretical and experimental data for NiO[13–23,34,48,49]. Our results have important implications for AF spintronic devices. Strong spin-phonon coupling provides a mechanism for optimizing spin wave propagation by tuning the phonon dispersion via geometrical confinement, stress or strain, and the knowledge of spin coupling to specific phonon polarizations (e.g. LO *vs.* TO) can be important for minimizing spin-wave damping or enhancing the spin-fluctuation mediated transmission of spin current.

**METHODS**

**Extraction of the spin – phonon coupling coefficient:** The spin-phonon coupling coefficient, $\lambda$, is calculated from the temperature-dependent UV Raman spectra. The frequency shift due to the spin-phonon coupling is given by $\Delta\omega_{sp-ph} = -\lambda S^2 \phi(t) = -\lambda \langle S_i \cdot S_j \rangle$, where $\phi_t$ is the short-range order parameter and $S_i$ represents the spin of Ni atom site *i*. As the temperature goes beyond Néel temperature, T$_N$, the short-range spin ordering $\langle S_i \cdot S_j \rangle$ quickly falls off. The calculation of $\phi_t$, using the mean field theory and the two-spin cluster theory, is already reported for S = 2 (FeF$_2$) and S = 2.5 (MnF$_2$)[25]. Besides, prior results of $\phi_t$ of S = 2 has been used as a resonable estimate of $\phi_t$ for S = 1 (Ni$^{2+}$ in NiF$_2$)[27], mainly because $\phi_t$ does not change much with S. In our case, we make use of $\phi_t$ for Ni$^{2+}$, to extract the spin-phonon coupling factor, $\lambda$, in NiO using the expression

$$\lambda = -\frac{\omega_R(T_{low}) - \omega_A(T_{low})}{[\phi(T_{low}) - \phi_-(2T_N)] \cdot S^2}$$

Here, $\omega_R(T_{low})$ is the observed Raman peak at the lowest temperature (T$_{low}$ ~ 80 K) and $\omega_A(T_{low})$ is the anharmonic prediction of that Raman peak at T = T$_{low}$. According to the above equation, the calculated spin-phonon coupling factors, $\lambda$, for 2TO and 2LO modes are -15.7 cm$^{-1}$ and 28.2 cm$^{-1}$. We divide the obtained $\lambda$ of 2TO and 2LO by two, to get the $\lambda$ for TO and LO mode, respectively. Since the second-order order peaks (2TO and 2LO) are more prominent in Raman spectra, we use them to extract $\lambda$ of TO and LO modes.

**Phonon band-structure calculation:** We used *ab initio* density functional theory (DFT) calculations, as implemented in the Vienna Ab-initio Simulation Package (VASP)[50,51], with the generalized gradient approximation (GGA) and the Perdew-Burke-Ernzerhof (PBE) parametrization for the exchange correlation functional. A plane wave basis set with kinetic energy cutoff of 550 meV is used to expand the electronic wave functions and an 8 × 8 × 8 Monkhorst Pack *k*-point mesh is adopted for the integration over the first Brillouin zone (BZ). The bulk NiO is relaxed until the forces were less than 0.001 eV/Å. The optimized lattice parameter of the unit cell is 4.179 Å, which is in good agreement with literature[17]. In case of magnetic NiO, collinear spin polarized calculations are performed for both relaxing the AF ground state and obtaining the phonon dispersion. To compensate the over-localization of the Ni 3*d* state, we make use of Dudarev DFT+U method[16,52], which combines DFT with a Hubbard Hamiltonian. As parameters of DFT+U (GGA+U), we used U = 8.0 eV and J = 0.95 eV from the literature[48], where U represents the energy increase for an extra on-site electron and J stands for the screened exchange energy[53]. Below the Néel temperature (523 K), the alternate (111) FM spin-sheets are slightly contracted, resulting in rhombohedral distortion[51]. We neglect this small rhombohedral distortion in the cubic NiO unit cell, since this structural distortion has very little effect on the phonon spectrum[16]. The phonon dispersion along the BZ path is calculated using a finite displacement scheme implemented in Phonopy[54], for a 2x2x2 supercell and 8x8x8 *k*-point mesh. The asymptotic long-range dipole-dipole interaction, which gives rise to LO-TO splitting, is included as a correction to the interatomic force constants by calculating the Born effective charge and dielectric tensor.

**Acknowledgements**

The work at UC Riverside was supported as part of the Spins and Heat in Nanoscale Electronic Systems (SHINES), an Energy Frontier Research Center funded by the U.S. Department of Energy, Office of Science, Basic Energy Sciences (BES) under Award # SC0012670. The *ab initio* simulations used the Extreme Science and Engineering Discovery Environment (XSEDE), which is supported by National Science Foundation (NSF) grant number ACI-1053575. M.M.L. acknowledges Conselho Nacional de Desenvolvimento a Pesquisa (CNPq) and the program Ciencias sem Fronteiras for financial support during her research at UC Riverside.

**Author contributions**

A.A.B. conceived the idea, coordinated the project and led the experimental data analysis; E.A., M.M.L., and F.K. conducted Raman spectroscopy, material characterization and experimental data analysis. B.D. performed *ab initio* computations and contributed to theoretical data analysis. R.K.L., Y.B. led theoretical data analysis. J.S. and J.X.L. contributed to the experimental and computational data analysis. All authors contributed to the manuscript preparation.

**FIGURE CAPTIONS**

**Figure 1: Raman spectroscopy data for NiO.** The room-temperature backscattering spectra are shown for (a) 488 nm and (b) 325 nm laser excitation. The labeled spheres indicate the position of the first-order and second-order Raman peaks. Temperature dependence of the frequencies of the second-order (c) 2TO and (d) 2LO Raman-active phonons in NiO. The spheres indicate all measured Raman peaks at different temperatures whereas the lines show the theoretical anharmonic trend for NiO fitted for the experimental data points above $T_N$. The difference of Raman peaks and anharmonic lines at low temperature is directly proportional to the spin-phonon coupling constant. Note that the spin-phonon coupling produces an opposite effect on the TO and LO phonon energies in NiO.

**Figure 2: Calculated phonon dispersion for NiO system with and without spin texture.** (a) Phonon dispersion of NiO with AF spin texture. The spheres show the position of the measured UV Raman peaks. The TO' mode is a zone-folded manifestation of TO-like mode at $k = (0.5,0,0.5)$. (b) Phonon dispersion of NiO with unpolarized spin. The red arrows show the frequency shift, $\Delta$, of the LO and TO phonon branches when AF spin ordering is included. (c) The AF primitive cell consisting of 2 Ni atoms and 2 O atoms and corresponding Brillouin zone used for the calculations of (a) and (b). The green arrows indicate the BZ path direction. (d) The unit cell resulting from doubling the AF primitive cell and corresponding Brillouin zone. (e) The phonon dispersion resulting from the AF unit cell in (d). The zone-folding of the TO' mode is apparent at $\Gamma$. Note a good agreement between the theory and experiment when the spin-phonon coupling is included.

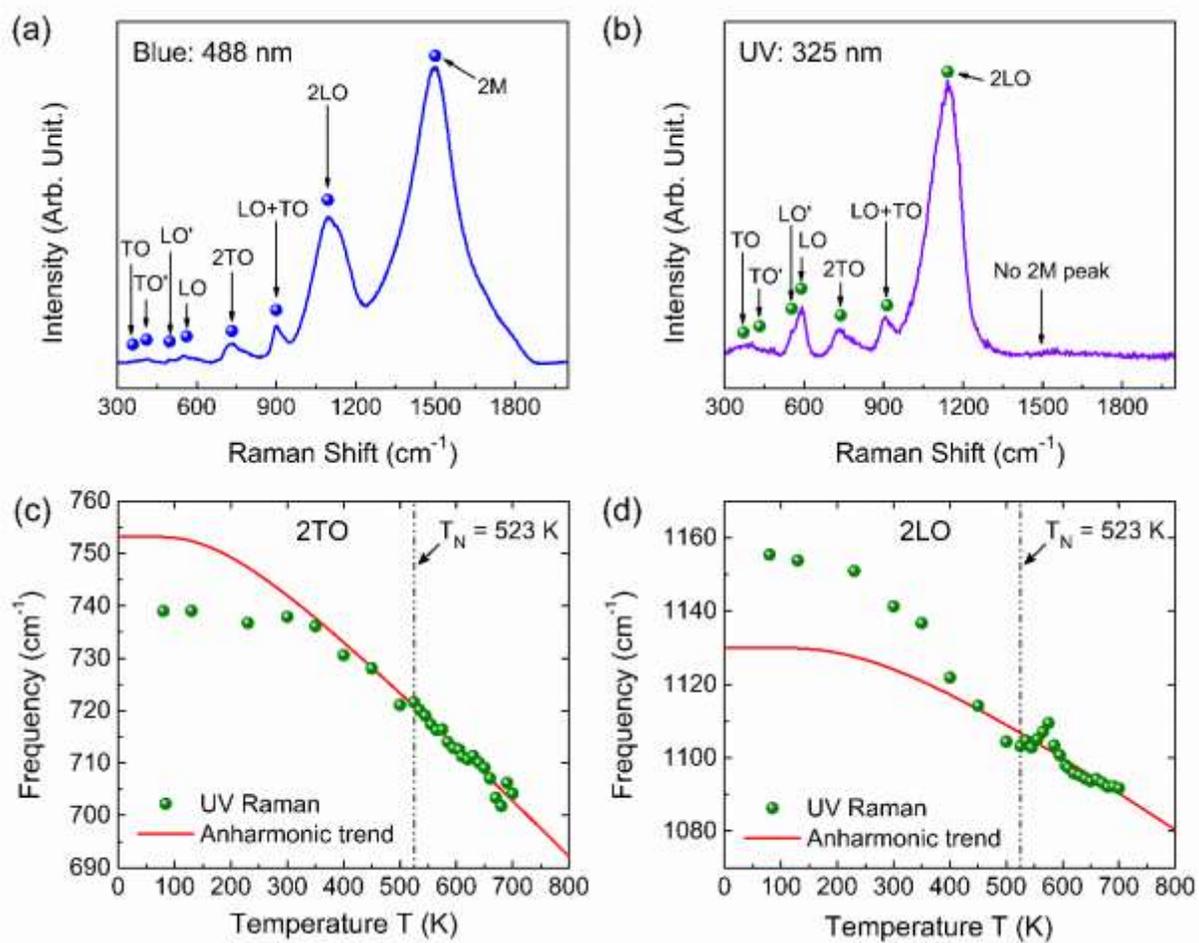

Figure 1

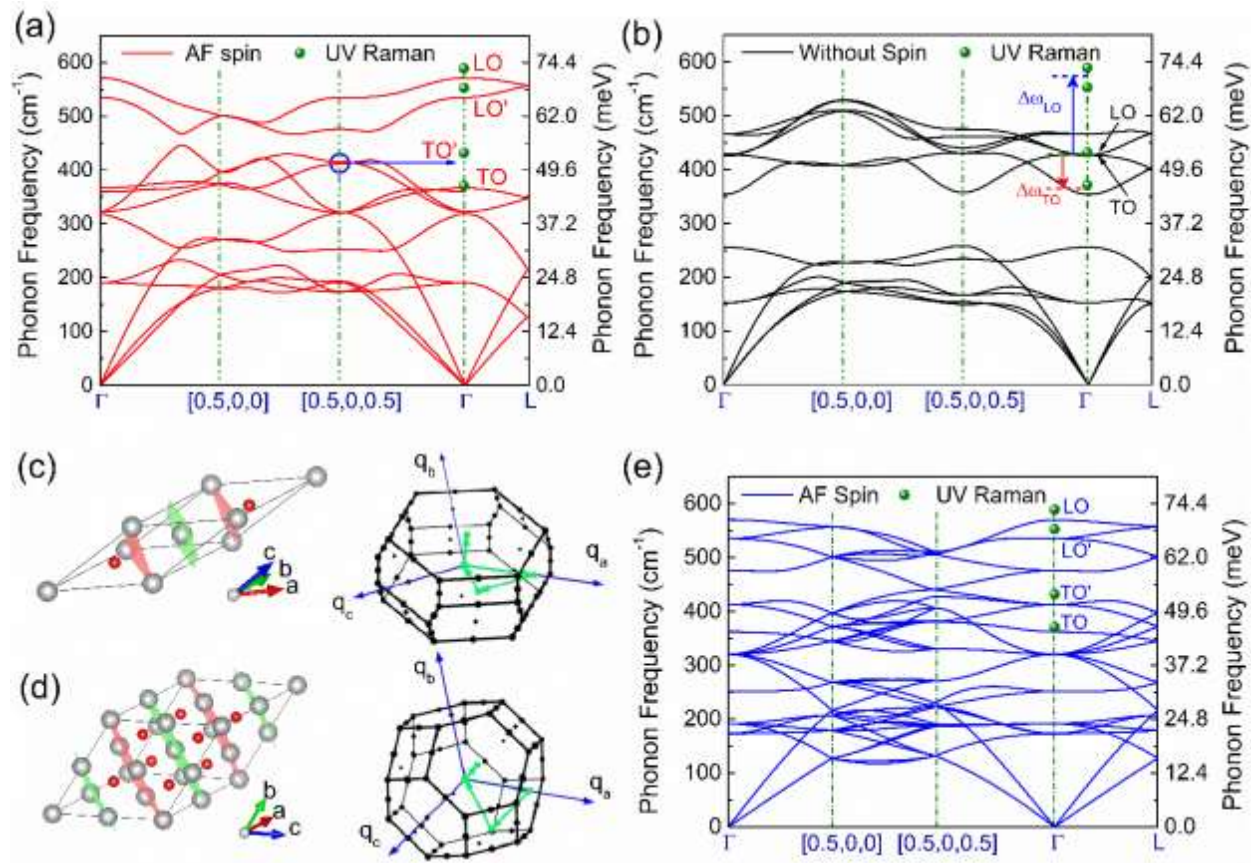

Figure 2